\documentclass{iau}
\usepackage{graphicx,natbib}
 
\newcommand{\apj}{ApJ}           
\newcommand{\apjl}{ApJ}           
\newcommand{\mnras}{MNRAS}       
\newcommand{\nat}{Nature}
\newcommand{\aap}{A\&A}

\newcommand{\aj}{AJ}
\newcommand{\pasp}{PASP}
\newcommand{\pasj}{PASJ}
\newcommand{\apjs}{ApJS}           

\newcommand{\msun}{\hbox{$M_\odot$}}
\newcommand{\re}{\hbox{$R_{\rm e}$}}

\newcommand{\reffig}[1]{Fig.~\ref{#1}}
\newcommand{\mlpop}{\hbox{$(M_\ast/L)_{\rm pop}$}}
\newcommand{\mldyn}{\hbox{$(M_\ast/L)_{\rm dyn}$}}
\newcommand{\atl}{ATLAS$^{\rm 3D}$}

\title[Scaling Relations of Early-Type Galaxies]{Dynamical Mass Determinations and Scaling Relations of Early-Type Galaxies}

\author[M. Cappellari]{Michele Cappellari$^1$}

\affiliation{$^1$Sub-department of Astrophysics, Department of Physics, University of Oxford, Denys Wilkinson Building, Keble Road, Oxford OX1 3RH. email: {\tt cappellari@astro.ox.ac.uk}}

\pubyear{2015}
\volume{311}
\jname{Galaxy Masses as Constraints of Formation Models}
\editors{M. Cappellari \& S. Courteau, eds.}

\begin{document}

\maketitle

\begin{abstract}
I review our understanding of classic dynamical scaling relations, relating luminosity, size and kinematics of early-type galaxies. Using unbiased determinations of galaxy mass profiles from stellar dynamical models, a simple picture has emerged in which scaling relations are driven by virial equilibrium, accompanied by a trend in the stellar mass-to-light ratio ($M/L$). This picture confirms the earliest insights. The trend is mainly due to the combined variation of age, metallicity and the stellar initial mass function (IMF). The systematic variations best correlate with the galaxy velocity dispersion, which traces the bulge mass fraction. This indicates a link between bulge growth and quenching of star formation. Dark matter is unimportant within the half-light radius, where the total mass profile is close to isothermal ($\rho\propto r^{-2}$).
\end{abstract}

\firstsection
\section{Introduction}

Dynamical scaling relations of early-type (elliptical E and lenticular S0) galaxies (ETGs) relate the size, luminosity ($L$) and stellar kinematics of galaxies. Sizes are typically described by the half-light radius (\re), while kinematics is generally quantified by the stellar velocity dispersion ($\sigma$) within a given aperture, which in this review I assume no larger than \re. Given that luminosity and size depend on distance, while kinematics do not, one of the first key applications of galaxy scaling relations was to infer galaxy distances \citep{Dressler1987,Djorgovski1987}.

In more recent times, more accurate distance determination techniques have been developed, mostly based on characteristics of the galaxy stellar population \citep[e.g.][]{Tonry2001}. Moreover we have a deeper understanding of the relation between redshift and distance \citep[e.g.][]{Planck2014}. For these reasons distance determination is not any more the main use of dynamical scaling relations.

Instead, dynamical scaling relations are nowadays a key tool to study galaxy formation. The main reasons for this is: (i) because they provide a statistical description for easily measurable characteristics of galaxies as a function of time (redshift), which can be directly compared with numerical simulations \citep[e.g.][]{Robertson2006,Boylan-Kolchin2006,Oser2012,Porter2014}; and (ii) due to the fact that scaling parameters are actually expected to evolve very differently depending on the galaxy formation mechanism \citep[e.g.][]{Naab2009,Hopkins2010}.

In this review I focus on what one can learn about dynamical scaling relations using dynamical models. Some of the results I describe are closely linked to findings made using strong lensing or stellar population synthesis approaches. These were reviewed separately at this Symposium, in particular by Tommaso Treu and Charlie Conroy, and are just briefly mentioned here. The reader is referred to \citet{Courteau2014} for a combined review of all these different mass determination techniques.

\section{Classic scaling relations}

\begin{figure}
\centering
\includegraphics[width=\columnwidth]{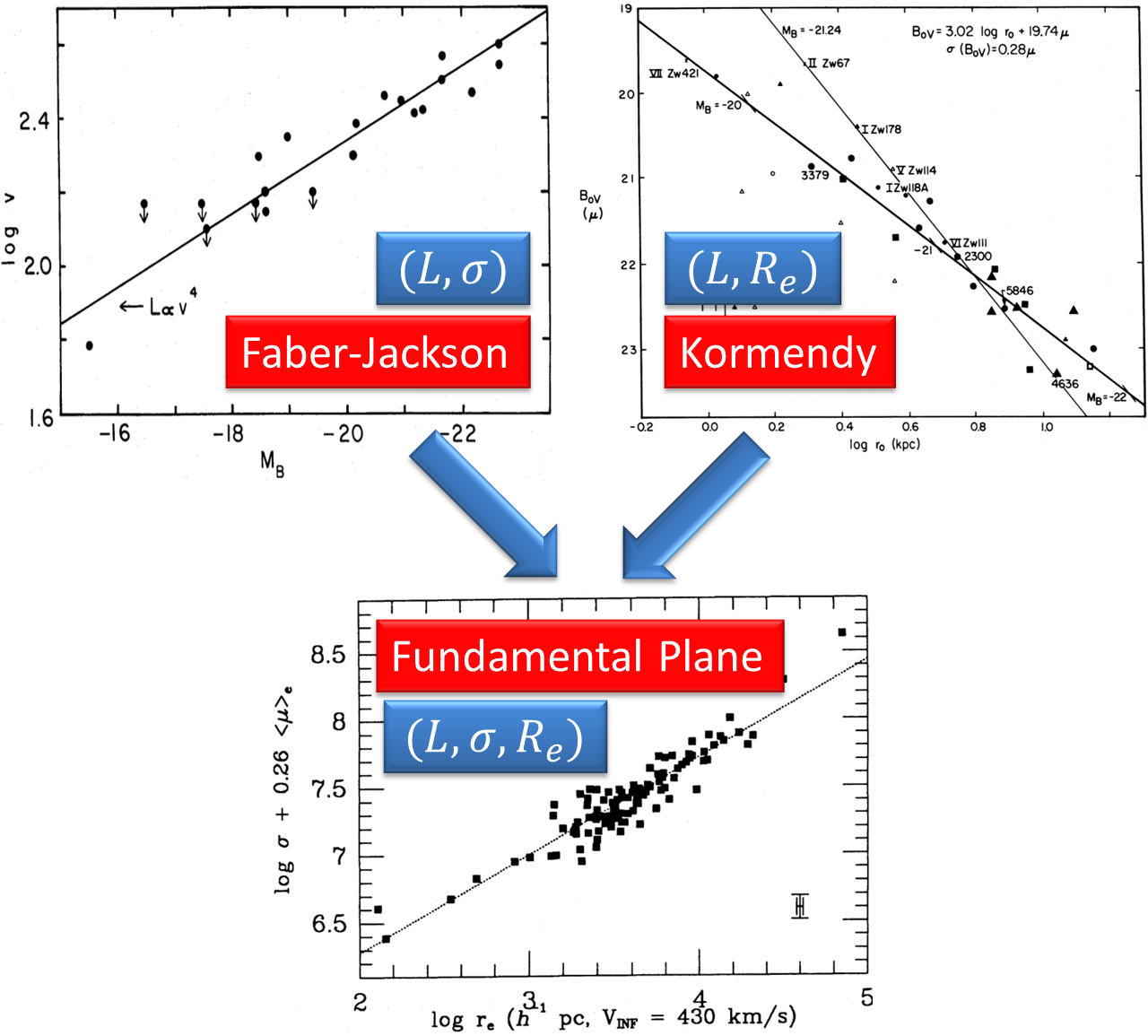} 
\caption{{\bf Classic scaling relations.} The Faber-Jackson and the Kormendy relations are two special projection of a more fundamental one, aptly named the Fundamental Plane. The three figures are taken from \citet{Faber1976}, \citet{Kormendy1977} and \citet{Djorgovski1987} respectively.}
\label{fig:classics}
\end{figure}

The first dynamical scaling relation to be discovered was the one between luminosity and stellar velocity dispersion (\citealt{Faber1976}, top-left in \reffig{fig:classics}). The observed relation had the form $L\propto\sigma^4$ and the authors pointed out it also suggests a trend between the mass-to-light ratio ($M/L$) and galaxy luminosity. 

Soon thereafter, a correlation between galaxy surface brightness $\Sigma$ and galaxy size was also found (\citealt{Kormendy1977}, top-right in \reffig{fig:classics}). When one defines the surface brightness as the mean value within \re, then $\Sigma_e=L/(2\pi R_e^2)$. This means that the Kormendy relation describes a correlation between galaxy radius and luminosity. This form has the advantage that it does not explicitly include galaxy size on both axes of the correlation. The $L-\re$ relation has recently become quite popular to study galaxy evolution as a function of redshift \citep[e.g.][]{vanderWel2014}, given that it does not involve any kinematic determination and for this reason is much more ``economical'' to observe than the Faber-Jackson.

Thanks to larger systematic surveys of ETGs it was later discovered that the Faber-Jackson and the Kormendy relations are just two special projections of a plane described by galaxies in $(\log L,\log\sigma,\log\re)$ coordinates \citep{Dressler1987,Djorgovski1987}. This plane was aptly named the Fundamental Plane (FP). It was found to hold for all ETGs, including S0s and E galaxies, with a scatter smaller than 20\% in \re\ \citep[e.g.][]{Jorgensen1996}.

The existence of the FP was interpreted as due to the fact that galaxies satisfy virial equilibrium $M\propto\sigma^2\re$, with $M$ the galaxy mass \citep{Faber1987}. However the exponents of the FP were found to deviate significantly from the virial predictions, a result confirmed by all numerous subsequent studies \citep[e.g.][]{Hudson1997,Scodeggio1998,Pahre1998,Colless2001,Bernardi2003fp}. In particular, a recent determination of the plane, for the \atl\ volume-limited survey \citep{Cappellari2011a}, with $\sigma_e$ measured from integral-field stellar kinematics within \re, gives $L\propto\sigma_e^{1.25} R_e^{0.96}$ (fig.~12 in \citealt{Cappellari2013p15}). This deviation of the FP from the virial predictions is called the ``tilt'' of the FP.

The original FP discovery papers suggested the variation in the stellar $M/L$ as a likely explanation for the tilt. In a brilliant proceedings paper, the Seven Samurai team went as far as stating that ``two-dimensionality implies that the virial theorem is the only tight constrain on E structure. (...) implies $(M/L)_e\sim L^{0.24\pm0.04}I_e^{0.00\pm0.06}$. Core and global $M/Ls$ agree well, implying that ellipticals are mainly baryon dominated within \re\ and that $M/Ls$ are stellar'' \citep{Faber1987}. As I will describe in what follows, it took a few decades for all these early insights to be convincingly confirmed. 

\section{Candidates for the tilt of the Fundamental Plane}

\begin{figure}
\centering
\begin{minipage}{.45\textwidth}
\includegraphics[width=\textwidth]{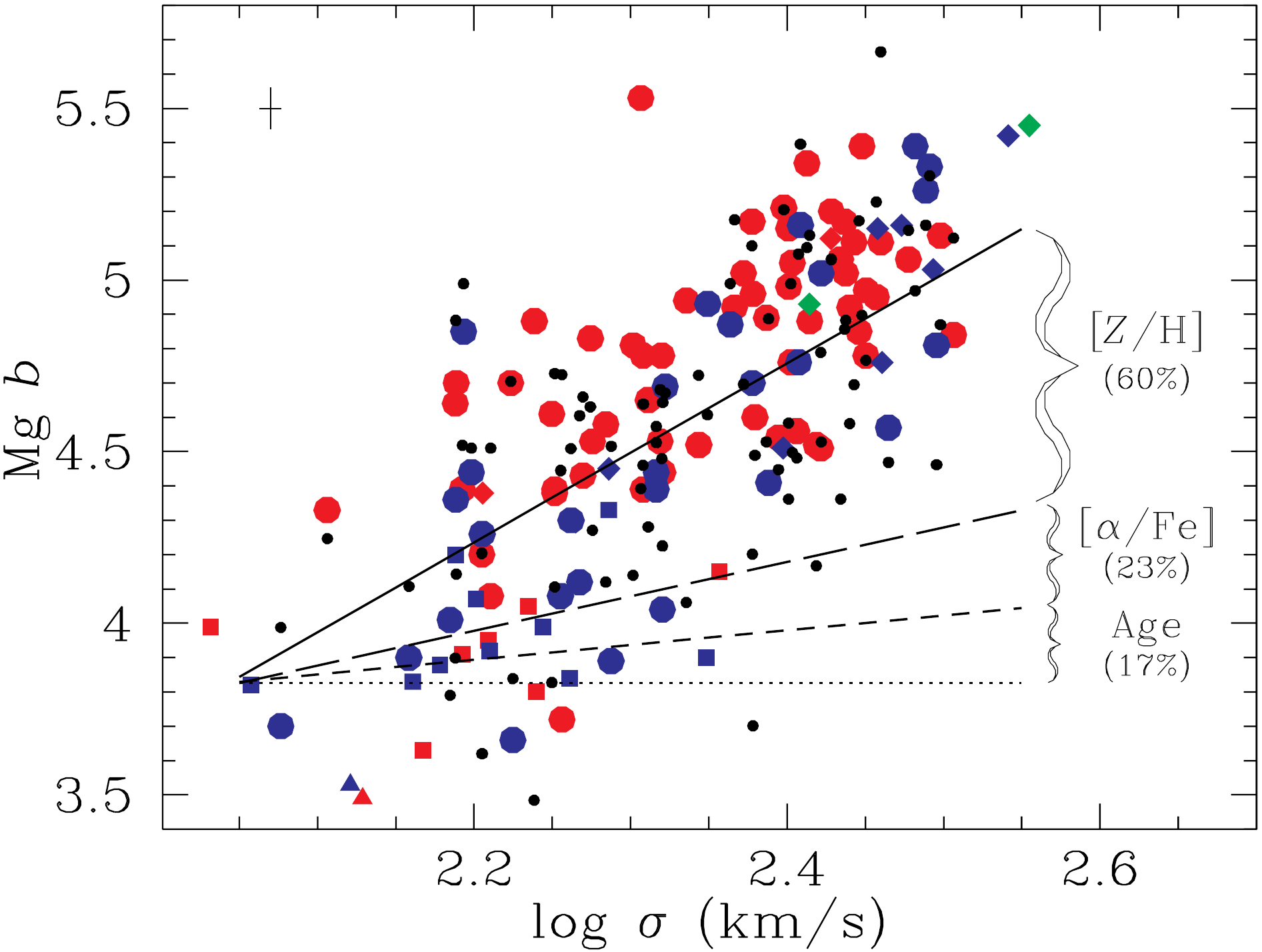}
\includegraphics[width=\textwidth]{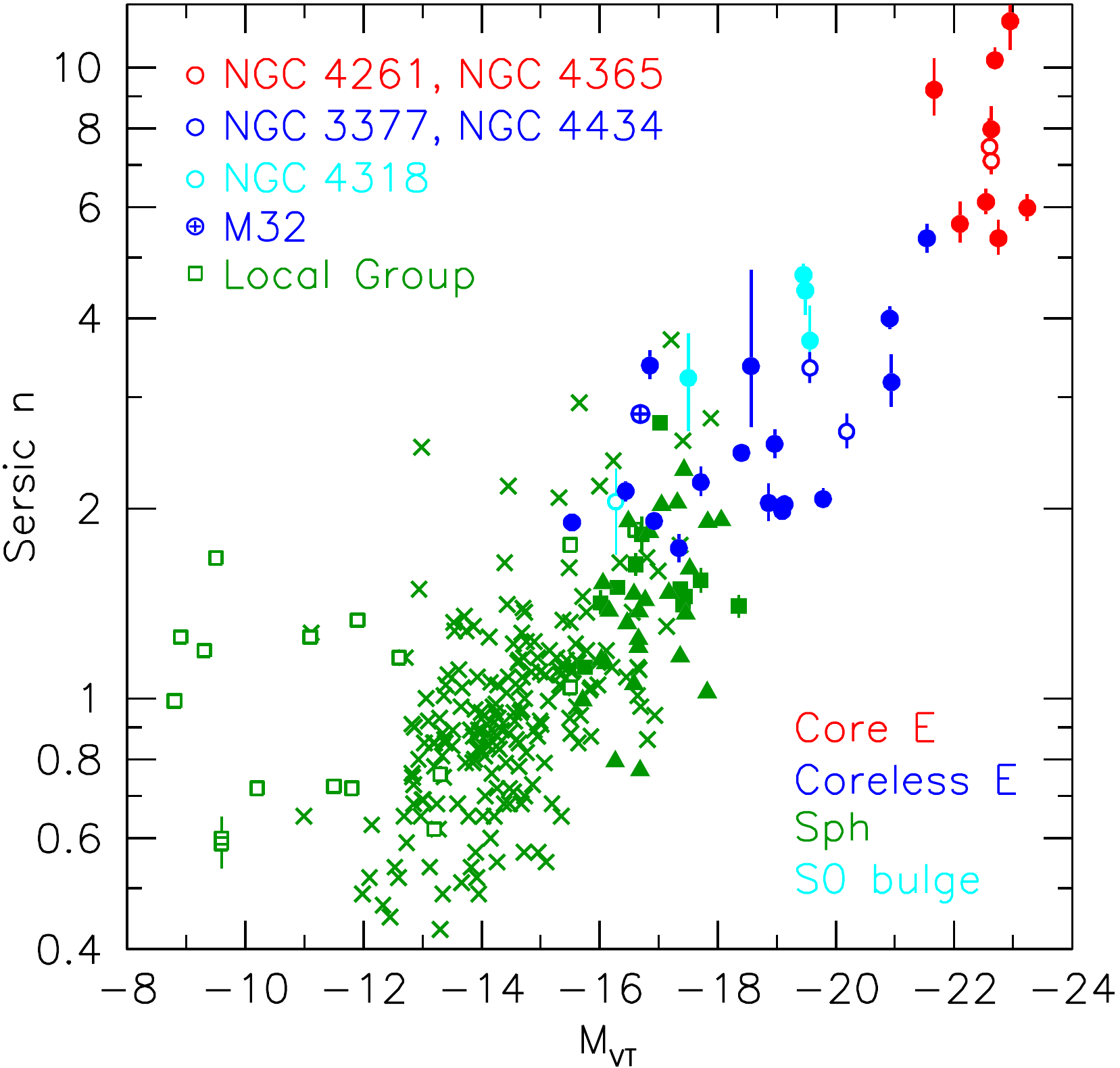}
\end{minipage}%
\begin{minipage}{.55\textwidth}
\includegraphics[width=\textwidth]{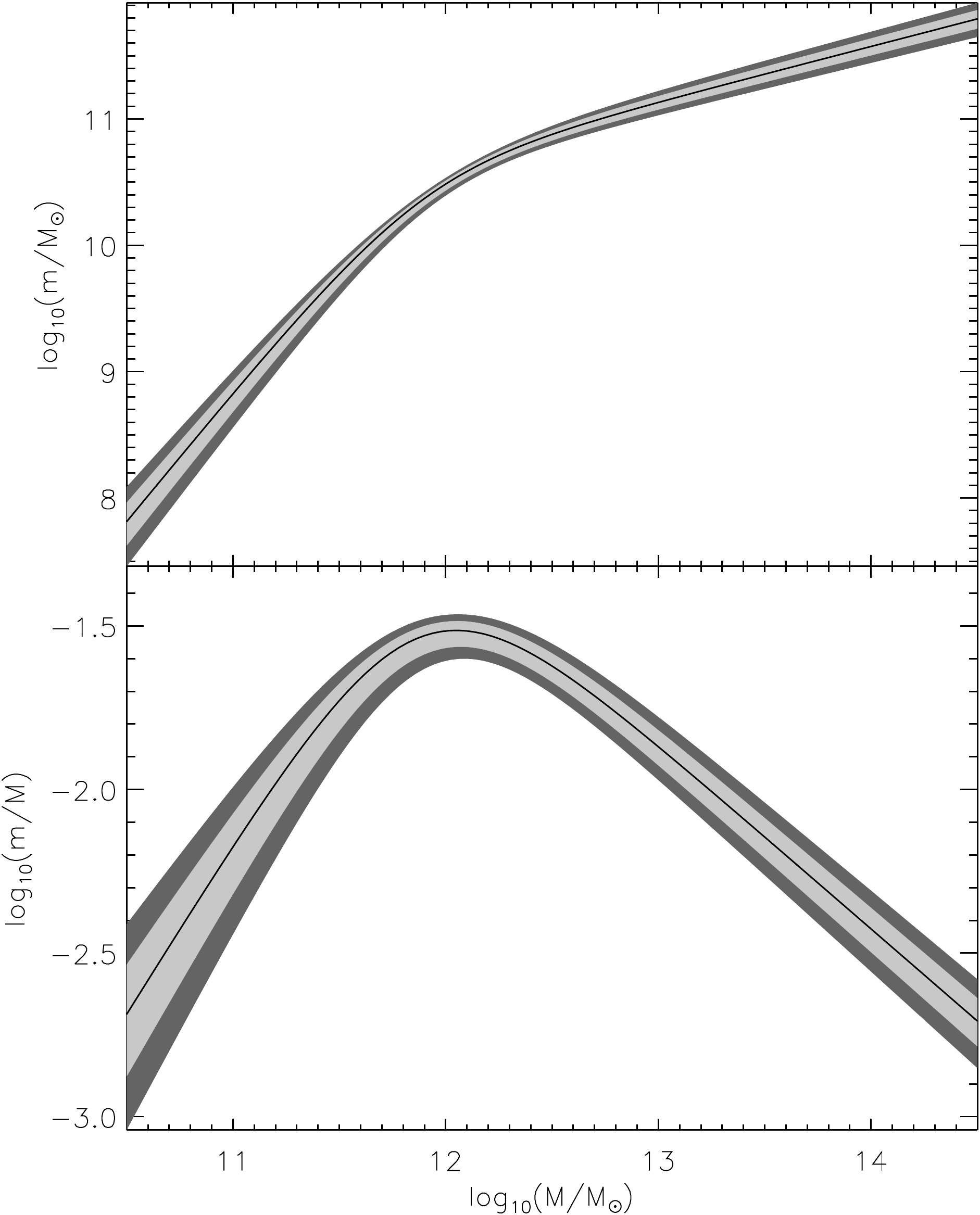}
\end{minipage}
\caption{{\bf Candidates for the FP tilt.} {\em Top-Left Panel:} Systematic variations in the galaxies stellar population (from \citealt{Thomas2005daniel}). {\em Bottom-Left Panel:} trends in the non-homology of the surface-brightness profile, as parametrized by the Sersic index (from \citealt{Kormendy2009}). {\em Right Panel:} Variations in the dark matter fraction (from \citealt{Moster2010}).}
\label{fig:tilt_candidates}
\end{figure}

The variation of the $M/L$ of the stellar population was immediately recognized as a potential source for the tilt of the FP. This is because systematic changes in the galaxy population were already known, with galaxies becoming older and more metal rich with increasing mass or $\sigma$ (e.g. \citealt{Thomas2005daniel}, \reffig{fig:tilt_candidates} top-left). This variation can potentially explain a major part of the FP tilt and scatter, predicting larger $M/L$ as a function of $\sigma$ by a factor of a few, as observed (depending on the photometric band) over the full range of galaxy masses \citep{Prugniel1996fp,Forbes1998fp}.

The surface brightness profiles of ETGs also display systematic variations as a function of their luminosity. The profiles become more concentrated, or have larger \citet{Sersic1968} indices, for increasing galaxy luminosity (\citealt{Caon1993}, \citealt{Graham2003}, \citealt{Kormendy2009}, bottom-left in \reffig{fig:tilt_candidates}). At fixed mass, a steeper profile implies a larger $\sigma$ within the central regions \citep{Ciotti1991} where the kinematics is observed (typically within a fraction of \re). The amount of $\sigma$ variation is again in principle sufficient to explain a major part of the FP tilt \citep{Ciotti1996,Graham1997fp,Prugniel1997fp,Bertin2002fp,Trujillo2004fp}.

A third potential cause for the FP tilt is the fraction of dark matter within the region where kinematics is observed. The dark matter fraction is expected to increase systematically with mass, for the range of interest of FP studies (e.g. \citealt{Moster2010}, \reffig{fig:tilt_candidates} right panel). This can cause variations in the observed total $M/L$ of an amount again sufficient to produce a significant fraction of the measured tilt \citep{Renzini1993,Borriello2003fp,Tortora2012}.

\section{Measuring unbiased $M/L$ in galaxies}

A way to remove most of the uncertainties in the debate about the source of the FP tilt of ETGs consists of accurately modelling stellar population, non-homology and dark matter in a quantitative way. This can be done using dynamical models of the stellar kinematics. Three main techniques have been used in the past decades: (i) \citet{Schwarzschild1979} numerical orbit-superposition technique; (ii) \citet{Syer1996} made-to-measure particle-based approach and (iii) \citet{Jeans1922} hydrodynamic equations.

\begin{figure}
\includegraphics[width=\columnwidth]{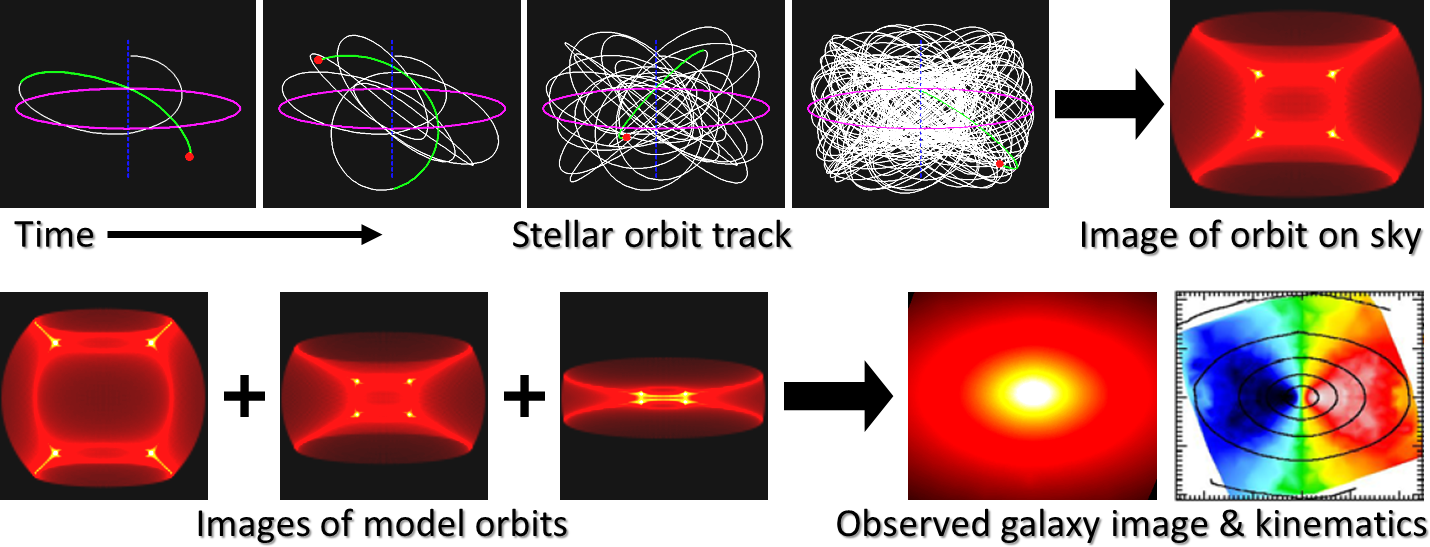}

\vspace*{0.5cm}
\includegraphics[width=\columnwidth]{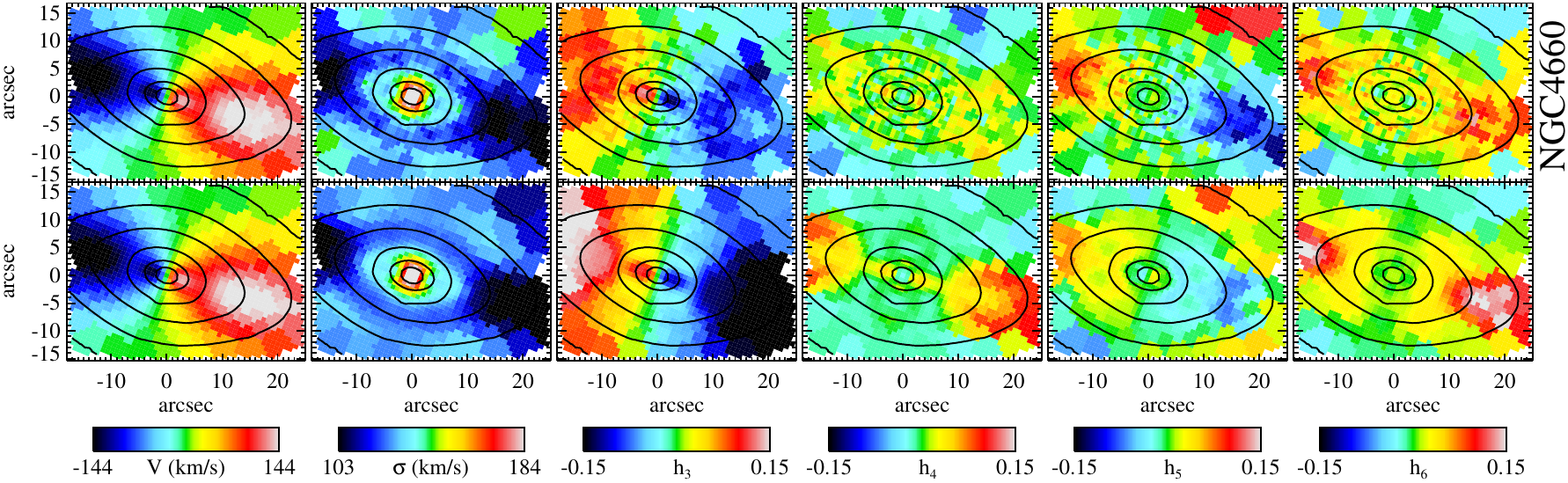} 
\caption{{\bf Schwarzschild's orbit-superposition method.} {\em Top Row:} numerical integration of a single orbit in the adopted gravitational potential. After a sufficiently long time the density (of regular orbits) converges to a fixed distribution. {\em Middle Row:} the method finds the linear combination of thousands of orbits (three representative are shown here) which best fits the galaxy image and stellar kinematics.
{\em Bottom two rows:} data (top) versus model (bottom) comparison. The model can fit the full stellar line-of-sight velocity distribution, here parametrized by the first six Gauss-Hermite moments (from \citealt{Cappellari2007}).}
\label{fig:orbits}
\end{figure}

The most popular has been \citet{Schwarzschild1979} numerical orbit-superposition technique \citep[e.g.][]{Richstone1988,Rix1997,vanDerMarel98,Gebhardt2003,Thomas2004,Cappellari2006,vanDenBosch2008}, which is able to find the general linear combination of model orbits which best fits the galaxy image and kinematics in exquisite detail (\reffig{fig:orbits}). A downside of the generality of the method is its lack of predictive power. As an example, one can fit axisymmetric Schwarzschild's models to a simulated edge-on barred galaxy. Even in this case, both the galaxy image and the kinematics will be reproduced in great detail. The good fit gives no indication that a bar is present and that the mass model and recovered orbital distribution are significantly in error. Systematic problems in the data can also be easily fitted without raising any concern. Similar strength and limitations are shared by the made-to-measure method \citep{deLorenzi07,Dehnen2009,Long2010}.

An alternative approach consists of solving the \citet{Jeans1922} equations of stellar hydrodynamics. Contrary to orbit or particle-based methods, with Jeans's method one has to make an assumption about the shape of the velocity ellipsoid in galaxies. Earlier papers assumed a semi-isotropic velocity ellipsoid ($\sigma_z=\sigma_R$), like in Jean's original work \citep[e.g][]{Satoh1980,Binney1990,vanderMarel1990,Emsellem1994,Magorrian1998}. But more recently, Schwarzschild's models based on multi-slit observations \citep[e.g.][]{Cappellari2002bh,Gebhardt2003,Thomas2009} and especially the constraining power of two-dimensional kinematics \citep{verolme02,Krajnovic2005,Cappellari2007}, have strongly excluded a semi-isotropic form for the velocity ellipsoid in ETGs, making these two-integral models outdated. 

However, the availability of good quality two-dimensional stellar kinematics for many galaxies \citep{Emsellem2004,Cappellari2011a} also revealed that galaxies have a relatively simple and predictable dynamics within $\sim1\re$. In fact, by making a simple generalization to Jeans's approach, removing the semi-isotropic assumption by allowing for $\sigma_z\neq\sigma_R$ \citep{Cappellari2008}, one can describe, or essentially ``predict'', the stellar kinematics of the large majority of real galaxies in surprisingly good detail, using just a couple of free parameters \citep{Scott2009,Cappellari2013p15}. An added advantage of this three-integral anisotropic Jeans approach is that it is orders of magnitude faster than the others. It allows for a quick exploration of parameters space for large samples of galaxies using a Bayesian statistical approach.

\section{Understanding scaling relations}

\begin{figure}
\includegraphics[width=\textwidth]{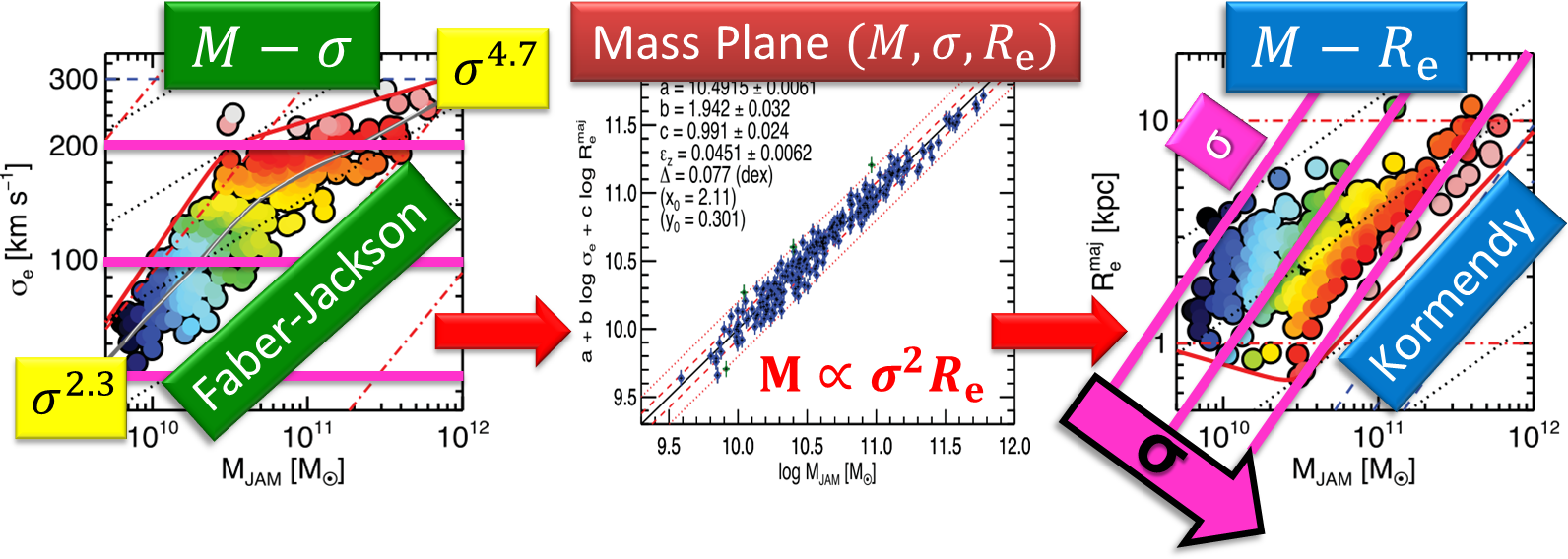}
\caption{{\bf Understanding scaling relations.} Replacing luminosity with dynamical mass, the Faber-Jackson shows a clean break at around $M\approx3\times10^{10}$ \msun\ ({\em Left Panel} from \citealt{Cappellari2013p20}). The FP becomes the Mass Plane, which follows the virial equation with high accuracy ({\em Middle Panel} from \citealt{Cappellari2013p15}). This implies that the $(M,\sigma)$ and the $(M,\re)$ projections contain the same information, apart from a change of coordinates. In both projections, galaxy properties best follow lines of constant $\sigma$ ({\em Right Panel}). In the left and right panels, the rainbow colours indicate the $M/L$.
}
\label{fig:mass_plane}
\end{figure}

Employing the semi-isotropic Jeans approach, and stellar kinematics of 37 galaxies from major/minor axis long-slit observations, it was found that the $M/L$ trend with galaxy mass remains nearly unchanged when one includes the effects of galaxy non-homology \citep{vanDerMarel91,Magorrian1998}. The model accuracy was improved by fitting models to two-dimensional stellar kinematics. Using a sample of 25 galaxies and both the Schwarzschild and Jeans approaches, \citet{Cappellari2006} found that the $(M/L)-\sigma$ relation is extremely tight and accounts for the entire scatter and tilt of the FP. In other words, when replacing luminosity with mass in the FP, the coefficients of the derived Mass Plane matched the virial predictions $M\propto\sigma^2\re$ within the errors. Independent confirmations were found by strong lensing studies \citep{Bolton2007mp,Auger2010}. This result was strengthened by modelling the 260 ETGs of the \atl\ sample with two-dimensional stellar kinematics. Galaxies were found to follow the virial predictions with high accuracy (\reffig{fig:mass_plane} middle). However, the study also pointed out the significant dependence of the plane coefficients on the technique used to measure them \citep{Cappellari2013p15}. This sensitivity can explain the apparent contrast between some of the past studies of the FP tilt.

Unbiased studies using detailed dynamical models found that the dynamically-derived \mldyn\ was related to the \mlpop\ inferred from stellar population models \citep{Gerhard2001}. This confirmed that at least part of the FP tilt is due to stellar population variations. It agrees with the fact that the scatter around the FP is also linked to variations in the stellar population \citep{Graves2009b,FalconBarroso2011,Springob2012,Magoulas2012}.

However, even improving the accuracy of the models using two-dimensional kinematics, the relation between dynamically-derived {\em total} $M/L$ and the stellar population \mlpop\ still showed significant systematic deviations \citep{Cappellari2006}. These could only be explained by either dark matter or IMF variations between galaxies. Deviations between accurate determinations of the stellar and total masses were also found using independent strong lensing techniques \citep[e.g.][]{Auger2010}.

\begin{figure}
\begin{minipage}{.33\textwidth}
    \includegraphics[width=0.95\textwidth]{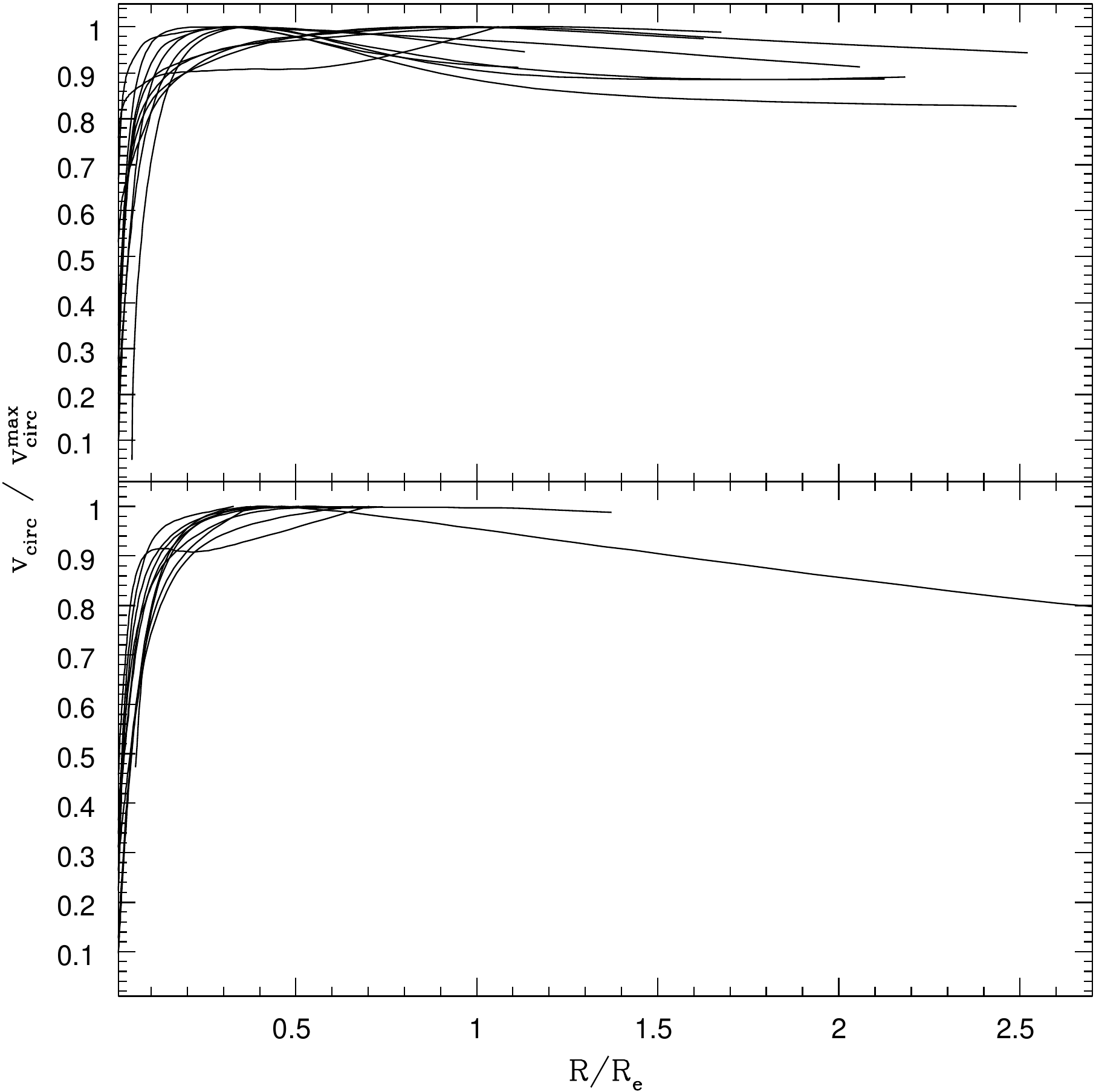}
\end{minipage}%
\begin{minipage}{.33\textwidth}
    \includegraphics[width=0.95\textwidth]{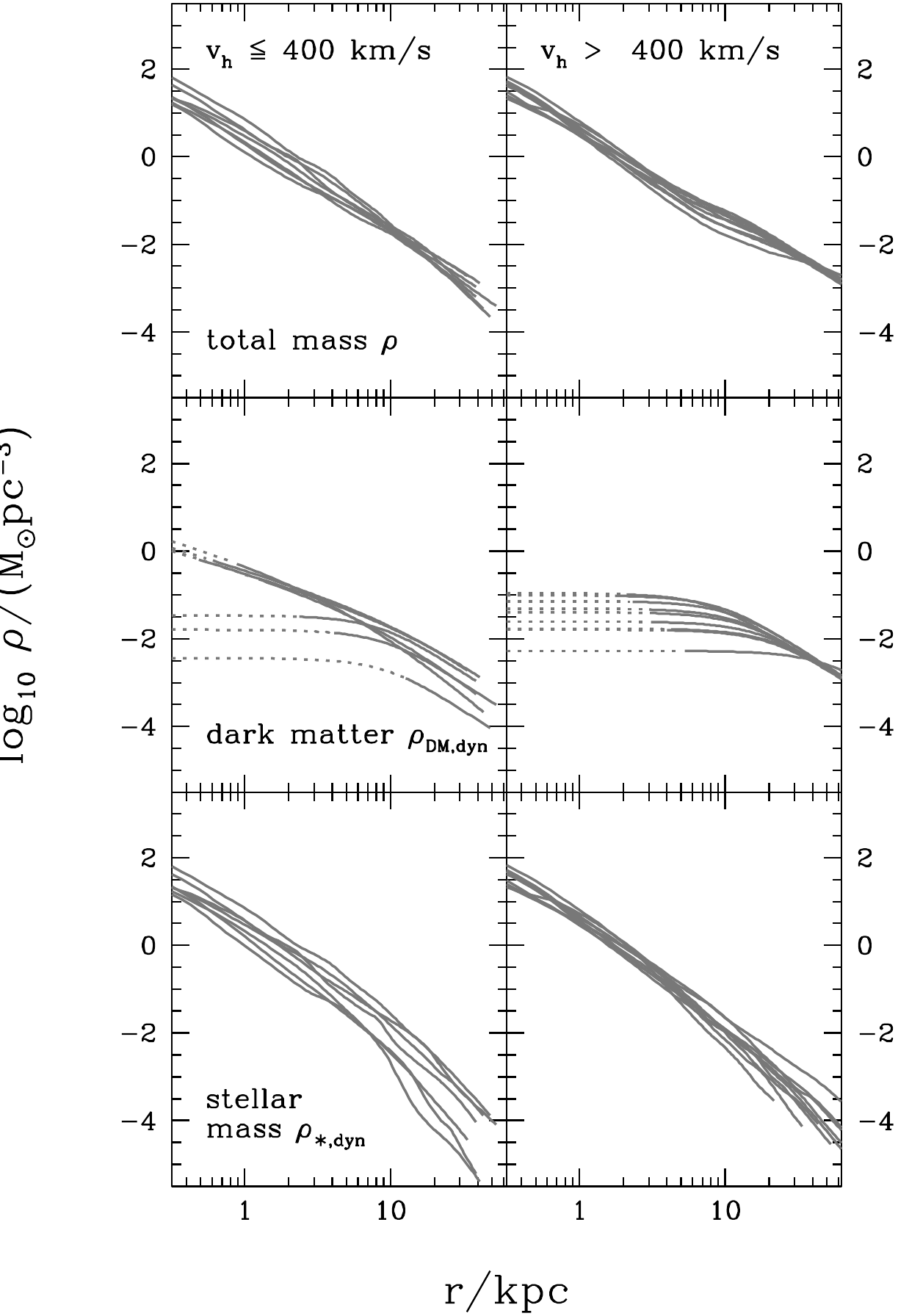}
\end{minipage}%
\begin{minipage}{0.33\textwidth}
    \includegraphics[width=0.95\textwidth]{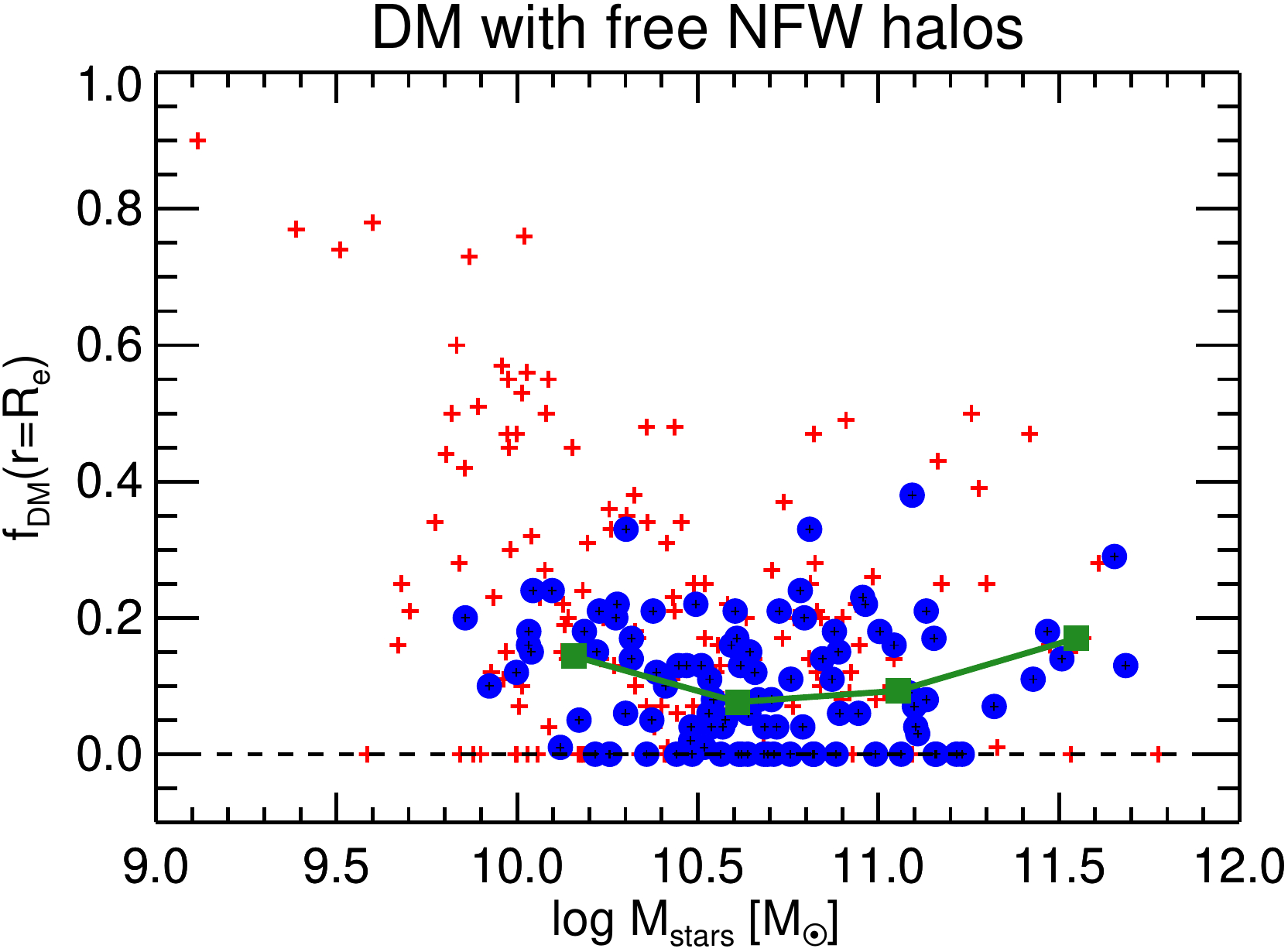} 
    \includegraphics[width=0.95\textwidth]{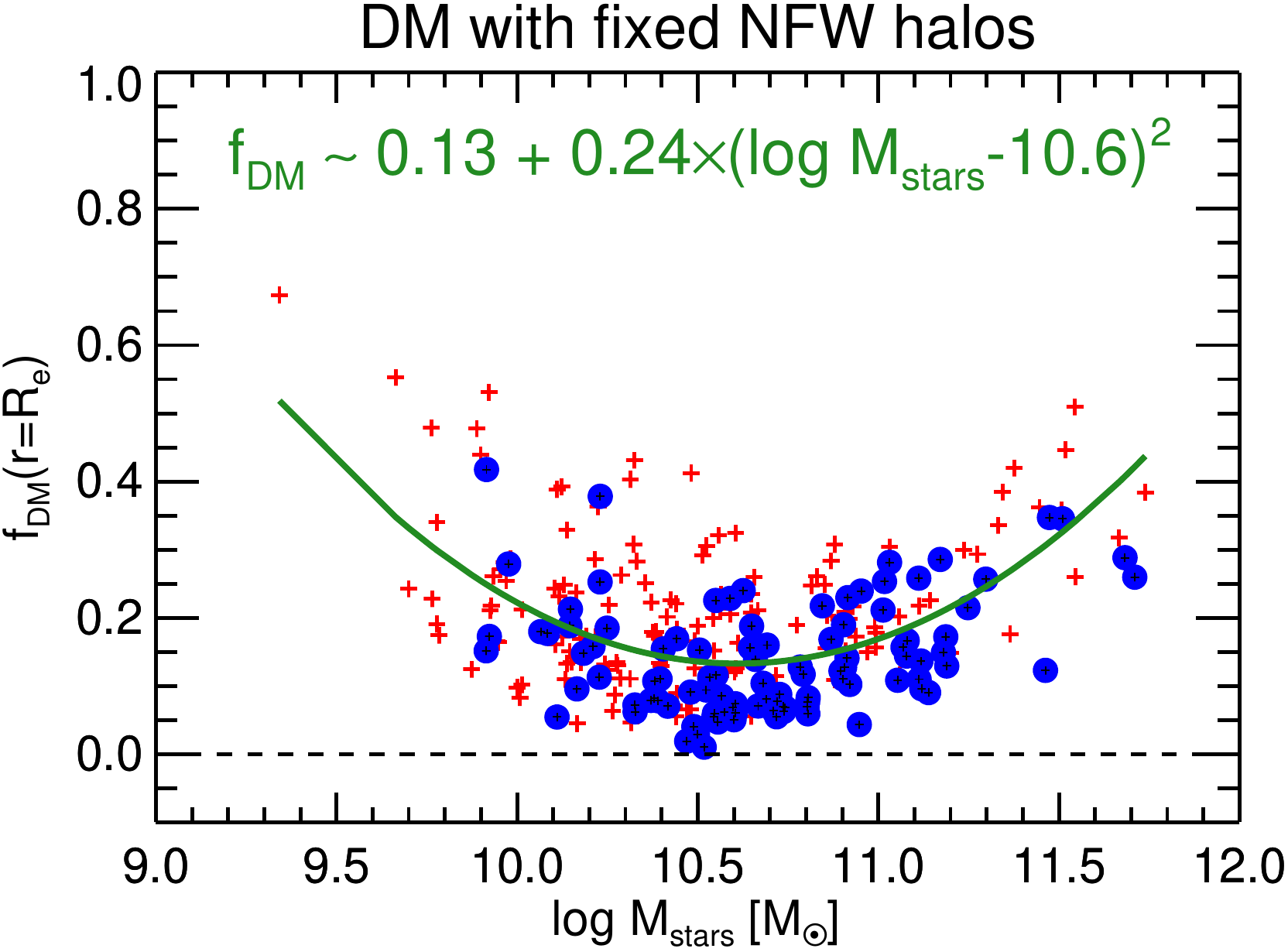} 
\end{minipage}
\caption{{\bf Dark matter from stellar dynamical models.} {\em Left Panels:} The circular velocity profiles of ETGs, as inferred from the models, are nearly flat within $\sim2\re$ (from \citealt{Gerhard2001}). {\em Middle Panels:} The total mass profiles are approximately isothermal ($\rho\propto r^{-2}$) within $\sim2\re$ (from \citealt{Thomas2011}). {\em Right Panels:} Either leaving the halo normalization as a free parameter in the models (top), or constraining it to the value predicted by $\Lambda$CDM, the inferred dark matter fractions within \re\  must be small, to fit the kinematics (from \citealt{Cappellari2013p15}).
}
\label{fig:dark_matter}
\end{figure}

To discriminate between dark matter and/or the IMF as the reason for the observed mass discrepancies, one needs to construct dynamical models which explicitly account for both the luminous and dark matter contributions. Unfortunately the problem is intrinsically quite degenerate. Still, using long-slit data and general models for two samples of about 20 galaxies, different studies appeared to agree that (i) dark matter represents a minor fraction of the total, within a sphere of radius $r\sim\re$; (ii) the {\em total} mass profile is nearly isothermal ($\rho\propto r^{-2}$) and produces almost flat circular rotation curves (\reffig{fig:dark_matter} left and middle), similarly to spiral galaxies. This was measured out to the median radius $r\sim2\re$ sampled by the kinematics \citep{Gerhard2001,Thomas2011} and in a few cases out to larger radii \citep[e.g.][]{Morganti2013}. The dark matter content within \re\ was better quantified with the modelling of the 260 ETGs of the \atl\ sample. A median dark matter fraction $f_{\rm DM}(r=\re)$ as low as 13\%  was measured for the full sample (\reffig{fig:dark_matter} right).

Thanks to the large sample and two-dimensional stellar kinematics, the \atl\ study was able to show that the systematic trend in the discrepancy between \mldyn\ and \mlpop\ could not be explained by a variation in the dark matter fraction. The most likely reason was then a systematic variation of the stellar initial mass function (IMF). This was inferred to vary in mass normalization from Milky-Way type \citep{Kroupa2001,Chabrier2003} to heavier than \citet{Salpeter1955} type, over the full mass range \citep{Cappellari2012}. This systematic trend was consistent with indications of a ``heavy'' IMF in massive ETGs from either stellar population \citep{vanDokkum2010} or strong gravitational lensing \citep{Auger2010imf}. This variation seems to be naturally explained by some theoretical models \citep[e.g.][]{Chabrier2014}.

\section{Implications for galaxy formation}

The fact that the FP is due to virial equilibrium implies that only a modest amount of information on galaxy formation is provided by the mere existence of the plane. Instead, most constraints on galaxy formation are encoded in the distribution of galaxy properties {\em within} the mass plane. A key empirical finding is that, within the plane, nearly all variation in galaxy properties is best described by a trend with the galaxy $\sigma$, rather than other global parameters like dynamical mass, size, surface brightness or \citep{Sersic1968} index (\reffig{fig:mass_plane}). This is true for the $M/L$ \citep{Cappellari2006}, stellar population indicators \citep{Graves2009b,Poggianti2013} as well as for the molecular gas fraction, colour and IMF \citep{Cappellari2013p20}.

\begin{figure}
\centering
\includegraphics[width=\textwidth]{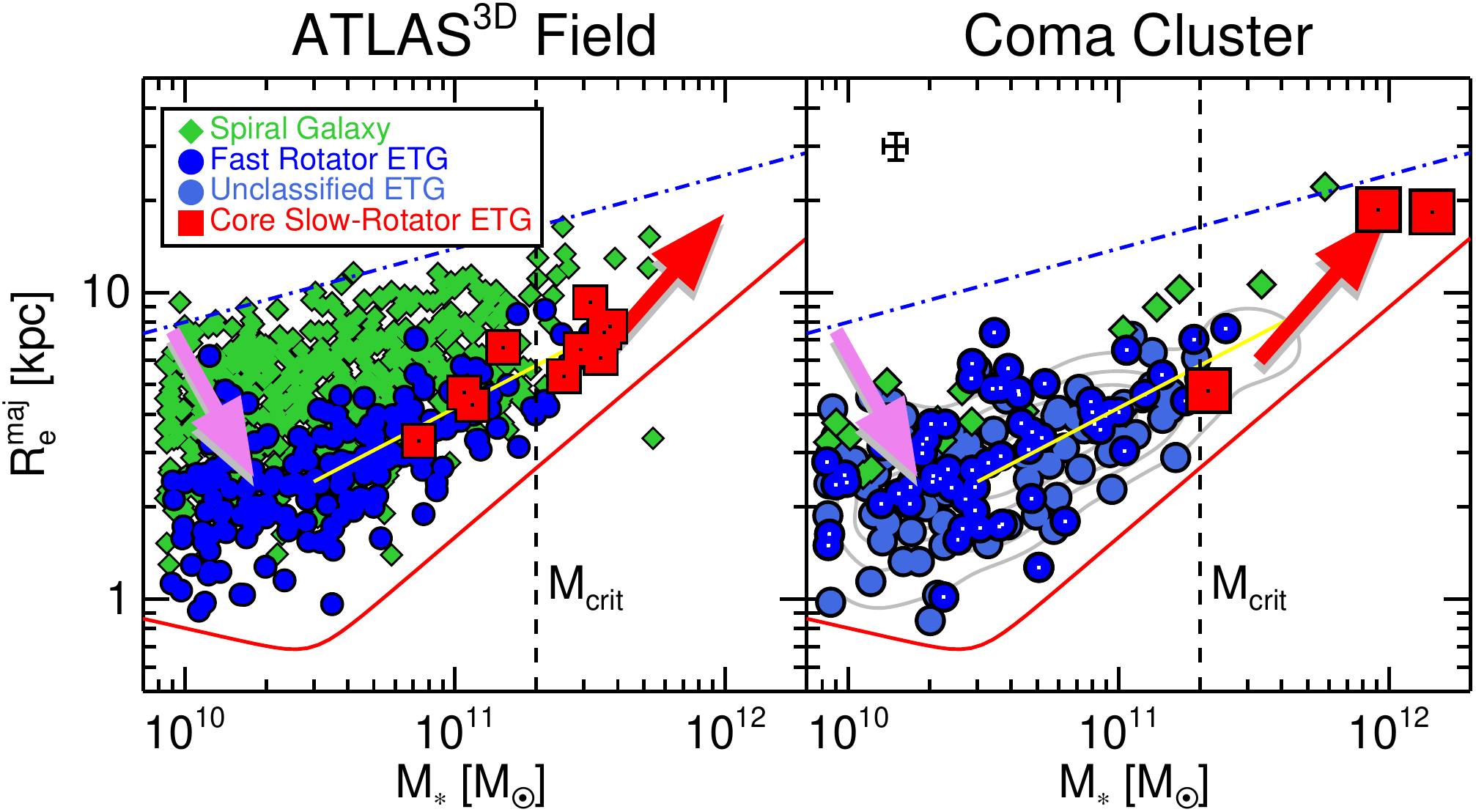}
\caption{{\bf The bulid-up of the Mass-Size relation.} The {\em Left Panel} shows a field sample. The {\em Right Panel} shows an identically-selected sample in one of the densest environments in the Universe. The magenta arrow qualitatively indicates the evolutionary track due to bulge growth and environmental quenching. In the dense environment, spirals are replaced by fast rotator ETGs, which have the same mass and size distribution as in the field sample. The red arrow shows the dry merging and halo quenching track, for slow rotators. These build up larger masses in denser environments (from \citealt{Cappellari2013apjl}).
}
\label{fig:spirals_etgs}
\end{figure}

It has become clear that $\sigma$ is a simple empirical tracer of the galaxy bulge {\em mass} fraction \citep{Cappellari2013p20}. And the correlations of galaxy properties with $\sigma$ describe a link between the mass growth in the central bulge and the cessation of the galaxy star formation. Similar results and conclusions are reached when $\sigma$ is replaced by a density measure within a fixed aperture \citep{Cheung2012,Fang2013}. This trend between bulge and quenching persists over the full range of galaxies morphological types, going smoothly from nearly bulge-less spiral galaxies, to the most dense disky-ellipticals fast-rotator ETGs (\reffig{fig:spirals_etgs}). However the bulge trend only exists below a characteristic mass $M_{\rm crit}\approx2\times10^{11}$ M$_\odot$. Above $M_{\rm crit}$ a different process is at work, which produces massive slow rotator ETGs with cores in their surface brightness (\reffig{fig:spirals_etgs}), preferentially lying near the centre of clusters or infalling groups \citep{Cappellari2013apjl,Fogarty2014}. 

The distinction between the first, bulge-related, quenching and the second, halo-related quenching, can be understood within the framework of hierarchical morphological evolution \citep{DeLucia2012}, where the massive slow rotating galaxies tend to form most of their mass efficiently at high redshift and generally remain the dominant galaxy of their own environment throughout the hierarchical assembly, starting from small groups to build more massive clusters. Slow rotators sink to the centre of mass of their groups by dynamical friction, where they further grow by dry merging, when the sub-groups merge. They are quenched when the halo reaches a sufficient mass to shock heat the infalling gas to the virial temperature, which prevents efficient cooling \citep{Birnboim2003,Keres2005}. Conversely, spiral galaxies form in a more gentle manner and build up their mass over a longer time. They tend not to dominate the mass of their group during the hierarchical growth. They swarm at large velocity around the cluster centre of mass, and are quenched by the cluster environment, with insignificant increase in their total masses (\reffig{fig:spirals_etgs}) while at the same time growing their bulges and becoming fast rotator ETGs \citep{Cappellari2013apjl}.

Defining or revisiting the above picture, is a challenge for the near future. This effort is currently driven by the synergy between large surveys of the distant Universe like CANDLES \citep{candles} and massive detailed studies of the nearby Universe using two-dimensional spectroscopy like SAMI \citep{Croom2012} and MaNGA \citep{Bundy2014}, combined with simulations of ever increasing spatial resolution and realism \citep[e.g.][]{Vogelsberger2014}.

\section*{Acknowledgements}

\noindent
I acknowledge support from a Royal Society University Research Fellowship.

\end{document}